\documentclass[conference,letterpaper]{IEEEtran}

\usepackage{comment}
\usepackage[utf8]{inputenc} 
\usepackage[T1]{fontenc}
\usepackage{url}
\usepackage{amssymb}
\usepackage{ifthen}
\usepackage{cite}
\usepackage[cmex10]{amsmath} % Use the [cmex10] option to ensure complicance
\newcommand\numeq[2]%
  {\stackrel{#1}{#2}}
  
\usepackage{epsfig,rotating,setspace,latexsym,amsmath,epsf,amssymb,amsfonts,bm,theorem,cite,algorithm,graphicx,epsf,authblk,epstopdf,color,algpseudocode,bbm,subcaption}

\usepackage{array}
\newcolumntype{P}[1]{>{\centering\arraybackslash}p{#1}}
\newcolumntype{M}[1]{>{\centering\arraybackslash}m{#1}}

\newtheorem{remark}{Remark}

\newtheorem{theorem}{Theorem}
\newtheorem{lemma}{Lemma}
\newenvironment{Proof}[1]{\medskip\par\noindent{\bf Proof:\,}\,#1}{{\mbox{\,$\blacksquare$}\par}}

\begin{document}
\title{Download Cost of Private Updating} 

% %%% Single author, or several authors with same affiliation:
\author[1]{Bryttany Herren}
\author[1]{Ahmed Arafa}
\author[2]{Karim Banawan}
\affil[1]{\normalsize Electrical and Computer Engineering Department, University of North Carolina at Charlotte, USA}
\affil[2]{\normalsize Department of Electrical Engineering, Alexandria University, Egypt}

\maketitle

%===
\begin{abstract}
We consider the problem of privately updating a message out of $K$ messages from $N$ replicated and non-colluding databases. In this problem, a user has an outdated version of the message $\hat{W}_\theta$ of length $L$ bits that differ from the current version $W_\theta$ in at most $f$ bits. The user needs to retrieve $W_\theta$ correctly using a private information retrieval (PIR) scheme with the least number of downloads without leaking any information about the message index $\theta$ to any individual database. To that end, we propose a novel achievable scheme based on \emph{syndrome decoding}. Specifically, the user downloads the syndrome corresponding to $W_\theta$, according to a linear block code with carefully designed parameters, using the optimal PIR scheme for messages with a length constraint. We derive lower and upper bounds for the optimal download cost that match if the term $\log_2\left(\sum_{i=0}^f \binom{L}{i}\right)$ is an integer. Our results imply that there is a significant reduction in the download cost if $f < \frac{L}{2}$ compared with downloading $W_\theta$ directly using classical PIR approaches without taking the correlation between $W_\theta$ and $\hat{W}_\theta$ into consideration.  
\end{abstract}

%===
\section{Introduction} \label{sec:introduction}
% PIR overview, Jafar original paper, general literature
The problem of private information retrieval (PIR), introduced by Chor et al. in \cite{ChorEtAl1998PIR}, seeks to find the most efficient way for a user to privately retrieve a single message from a set of $K$ messages from $N$ fully replicated and non-communicating databases.  PIR schemes are designed to download a \emph{mixture} of all $K$ messages, with the least number of overhead downloaded bits, such that no single database can infer the identity of the desired message. The user accomplishes this task by sending a query to each database. The databases respond truthfully to the submitted query with an answer string. The user can reconstruct the desired message from jointly \emph{decoding} the returned answer strings. Recently, the problem of PIR has received a growing interest from the information and coding theory communities. The classical PIR problem is re-formulated using information-theoretic measures in the seminal work of Sun-Jafar \cite{SunJafar2017CapacityOfPIR}. In there, the performance metric of the PIR scheme is the retrieval rate, which is the ratio of the number of the desired message symbols to the total number of downloaded bits. The supremum of this ratio is denoted by the PIR capacity, $C$. Sun and Jafar characterize the PIR capacity of the classical PIR model to be
\begin{align}\label{eq:pirC}
    C= \left ( 1 + \frac{1}{N} + \frac{1}{N^2} +\cdots + \frac{1}{N^{K-1}} \right )^{-1}.
\end{align}
Following \cite{SunJafar2017CapacityOfPIR}, the capacity (or its reciprocal, the normalized download cost) of many variations of the problem have been investigated, see, e.g., \cite{Banawan_coded, SPIR, MM-PIR, arbitraryCollusion, WangSkoglund, TianSun_upload, ChaoTian_leakage, Banawan_Byzantine, AttiaTandonStorage,PrivateComputation, PIR_cache_edge,Tamo_journal,Byzantine-Kang,LeakyPIR,OneShot_Rouhayeb}.

%Motivation: model updating ML, you do not need to download the entire msg, questions
In all these works, the user is assumed to have no information about the desired message prior to retrieval. Thus, the queries are designed independently from the message contents. This is not always the case in practice. To see that, consider the following classical motivational example of PIR: in the stock market, investors need to privately retrieve some of the stock records, since showing interest in a specific record may undesirably affect its value. PIR is a natural solution to this problem. Now, consider the case when an investor has already retrieved a specific stock record some time ago but this record has been changed. The investor needs to update the record at his/her side. A trivial solution to this problem is to re-apply the original PIR scheme again. Nevertheless, this solution overlooks the fact that stock records are \emph{correlated} in time. Another example arises in the context of private federated submodel learning \cite{JafarPrivateSubmodelFederated}, in which a user needs to retrieve the up-to-date desired submodel without leaking any information about its identity. The weights of each submodel are usually correlated in time as in the stock market example. In both examples, it is interesting to investigate whether or not the investor (user) can exploit the correlation between the outdated record (submodel) and its up-to-date counterpart to drive down the download cost. In this work, we focus our attention on a specific type of correlation, in which the up-to-date message is a distorted version of the outdated message according to a \emph{Hamming distortion} measure. The most closely related works to this problem are the PIR problems with side information, e.g.,  \cite{ChenWangJafarPIR-SI,WeiBanawanUlukus2019PIRPartiallyKnownSI,StorageConstrainedPIR_Wei,MMPIR_PSI,SSMMPIR_SI1,SSMMPIR_SI2,kadhe2017private}. In all these works, the user has side information in the form of a subset of \emph{undesired} messages, which are utilized to assist in privately retrieving the desired message. This is different from our setting, in which the user possesses side information in the form of an outdated \emph{desired} message. Furthermore, these works differ from each other in whether the privacy of the side information should be maintained or not. This is different from our problem in which the identity of the desired and side information is the same, and therefore the privacy constraint in our problem is modified to reflect this fact.

% Problem Formulation
In this paper, we introduce the problem of {\it private updating} for a message out of a $K$-message library from $N$ replicated and non-colluding databases. In this problem, the user has an \emph{outdated} version of the desired message $\hat{W}_\theta$, and wishes to update it to its up-to-date version $W_\theta$. Furthermore, the user has information about the {\it maximum} Hamming distance $f$ between the up-to-date message and its outdated counterpart, i.e., the user possesses $\hat{W}_\theta$, which differs in {\it at most} $f$ bits from the desired up-to-date message $W_\theta$. Based on $\hat{W}_\theta$ and $f$, the user needs to design a query set to reliably and privately decode the up-to-date version of the desired message $W_\theta$ with the least number of downloaded bits. Equivalently, the user needs to privately retrieve an \emph{auxiliary} message that corresponds to the flipped bit positions in the desired message. Similar to the works of \cite{PrivateSearch, PSI}, we assume that the databases can construct a \emph{mapping} from the original library of messages into a more appropriate form that can assist the user in the retrieval process. We aim at characterizing the optimal download cost needed to update $\hat{W}_\theta$ to $W_{\theta}$ without disclosing the desired message index $\theta$ to any of the databases.  

% Contributions
To that end, we propose a novel achievable scheme that is based on the \emph{syndrome decoding} idea introduced in \cite{PradhanRamchandran2003DISCUS}, and adapt it to our setting to exploit the correlation between $W_\theta$ and $\hat{W}_\theta$. Hence, syndrome decoding is used to \emph{compress} the desired message based on the user's side information (i.e., the outdated message $\hat{W}_{\theta}$). More specifically, the databases apply a linear transformation to the stored library of messages using the parity check matrix of a linear block code with carefully chosen parameters. The existence of such a code can be readily inferred from the Gilbert-Varshamov and the Hamming bounds \cite{blahut2003algebraic}. This transformation, in effect, maps the messages into their corresponding syndromes. Thus, the problem is reduced to retrieving the auxiliary messages (i.e., the syndrome representation) that comprises of $\left\lceil\bar{L}\right\rceil =\left\lceil\log_2 \left (\sum_{i=0}^{f} \binom{L}{i} \right )\right\rceil \leq L$ bits, where $L$ is the original message length. This enables us to directly apply the PIR scheme in \cite{SunJafar2017ArbitraryMessageLengthPIR} to the auxiliary messages of length $\left\lceil\bar{L}\right\rceil$, which is optimal under message length constraints. We confirm the validity of our proposed scheme by deriving a matching converse proof. Our converse proof is inspired by the converse proofs of the PIR problem with side information in \cite{WeiBanawanUlukus2019PIRPartiallyKnownSI,ChenWangJafarPIR-SI}, with the main difference being the fact that the side information in our case is the outdated message $\hat{W}_\theta$ in contrast to the cached messages. Consequently, we show that the optimal download cost, $\bar{D}_L$, is bounded by $\left\lceil\frac{\bar{L}}{C}\right\rceil\leq\bar{D}_L\leq\left\lceil\frac{\lceil\bar{L}\rceil}{C}\right\rceil$. Our achievable scheme is optimal if $\bar{L}$ is an integer, otherwise the gap between the upper and lower bounds is upper bounded by 2 bits. This justifies the efficacy of using syndromes as a message mixing technique in our setting. Furthermore, our results show that performing direct PIR on the original library of messages is strictly sub-optimal as long as the maximum Hamming distance $f < \frac{L}{2}$.

%===
\section{System Model} \label{sec:systemmodel}

We consider a classical PIR problem with $K$ independent, uncoded, messages $W_1, \cdots, W_K$, with each message consisting of $L$ independent and uniformly distributed bits. We have
\begin{align}
    H(W_i) &= L,\quad 1\leq i\leq K, \label{eq:Wlength} \\
    H(W_1, \cdots, W_K) &= H(W_1) + \cdots + H(W_K). \label{eq:Windep}
\end{align}
The $K$ messages are stored in $N$ replicated and non-communicating databases. The user (retriever) has a local copy of one of the messages whose index $\theta\in[K]$ is known to the user,\footnote{$[K]$ denotes the set $\{1,2,\dots,K\}$.} but not the database.\footnote{This is true if message $\theta$ has been previously obtained in a private manner.}
However, this message stored locally is {\it outdated}, and the user wishes to update it so that it is consistent with the copies in the databases without revealing to any of the databases what the message index is.
This setting defines the {\it private updating problem}.

Since each message is a string of $L$ bits, the problem can be formulated as privately determining which subset of the message bits need to be flipped in order to fully update it.
To model this, we use $\hat{W}_\theta$ to represent the locally stored outdated message, $\bar{W}_\theta$ to represent the subset of bit indices that need to be flipped, and $f$ to represent the {\it maximum} Hamming distance between $W_\theta$ and $\hat{W}_\theta$. Therefore, in order to update message $\theta$ the user needs to flip {\it at most} $f$ bits, i.e., $\bar{W}_\theta$ takes a value out of 
$\sum_{i=0}^{f} {L \choose i}$ choices. We assume that such choices are uniformly distributed and independently realized from $\hat{W}_\theta$. Based on this model, the following holds:
\begin{align}
    H(W_\theta) = H(\hat{W}_\theta) & = L, \label{eq:WhatLength} \\
    H(\bar{W}_\theta) = \log_2{ \left (\sum_{i=0}^{f} {L \choose i} \right )} & \triangleq \bar{L}, \label{eq:WbarLength} \\
    H(W_\theta | \hat{W}_\theta) = H(\bar{W}_\theta | \hat{W}_\theta) & = \bar{L}, \label{eq:WbarIndep} \\
    H(\bar{W}_\theta | \hat{W}_\theta, W_\theta) & = 0, \label{eq:WbarJoint} \\
    |\bar{W}_\theta| \leq f & \leq L \label{eq:frange},
\end{align}
where $|\cdot|$ denotes cardinality. For the purposes of this paper, we assume that the maximum Hamming distance $f$ between the outdated and updated message is known to the user. 

In order to retrieve $W_\theta$, the user sends a set of queries $Q^{[\hat{W}_\theta, f]}_1, \dots, Q^{[\hat{W}_\theta, f]}_N$ to the $N$ databases to efficiently obtain $\bar{W}_\theta$.
The queries are generated according to $\hat{W}_\theta$ and $f$, and are jointly independent of the realizations of the $[K]\backslash\{\theta\}$ messages and $\bar{W}_\theta$ given $\hat{W}_\theta$.
Therefore we have\footnote{We use the notation $x_S$ to denote the collection of $\{x_i,~i\in S\}$.}
\begin{align}
    I\left(W_{[K]\backslash\{\theta\}}, \bar{W}_\theta ; Q^{[\hat{W}_\theta, f]}_{1:N} \Big| \hat{W}_\theta \right ) =0. \label{eq:Qindep}
    % \\ I\left(W_{[K]\backslash\{\theta\}}, \bar{W}_\theta ; Q^{[\hat{W}_\theta, f]}_1, \cdots, Q^{[\hat{W}_\theta, f]}_N\right) = 0
\end{align}
Upon receiving the query $Q^{[\hat{W}_\theta, f]}_n$, the $n$th database replies with an answering string $A^{[\hat{W}_\theta, f]}_n$, which is a function of $Q^{[\hat{W}_\theta, f]}_n$ and all the $K$ messages stored. Therefore, $\forall\theta \in [K],~ \forall n \in [N]$, we have
\begin{align}
    H\left(A^{[\hat{W}_\theta, f]}_n \Big| Q^{[\hat{W}_\theta, f]}_n, W_{1:K}\right) = 0. \label{eq:Afunc}
\end{align}

To ensure that individual databases do not know which message is being updated, we need to satisfy the following {\it privacy constraint}, $\forall n \in [N],~ \forall k \in [K]$:
\begin{align}
    \!\left(\!Q^{[\hat{W}_1, f]}_n, A^{[\hat{W}_1, f]}_n, \hat{W}_1, W_{1:K}\!\right) \!\!\sim\!\! \left(\!Q^{[\hat{W}_k, f]}_n, A^{[\hat{W}_k, f]}_n, \hat{W}_k, W_{1:K}\!\right)\!, \label{eq:privacy}
\end{align}
where $\sim$ denotes statistical equivalence. After receiving  the answering strings $A^{[\hat{W}_\theta, f]}_{1:N}$ from all the $N$ databases, the user needs to decode the desired information $W_\theta$ with no uncertainty, satisfying the following {\it correctness constraint}:
\begin{align}
    H\left(W_\theta \Big| A^{[\hat{W}_\theta, f]}_{1:N}, Q^{[\hat{W}_\theta, f]}_{1:N}, \hat{W}_\theta\right) = 0. \label{eq:correct}
\end{align}

For fixed $N$, $K$, and $f$, a pair $(\bar{D}, L )$ is {\it achievable} if there exists a private updating scheme for messages of length $L$ bits long satisfying the privacy constraint \eqref{eq:privacy} and the correctness constraint \eqref{eq:correct}.
In this pair, $\bar{D}$ represents the expected number of downloaded bits received from the $N$ databases independently via the answering strings $A^{[\hat{W}_k, f]}_{1:N}$, i.e., 
\begin{align}
    \bar{D} = \sum_{n=1}^{N} H \left ( A^{[\hat{W}_\theta, f]}_n \right ). \label{eq:generalDLcost}
\end{align}
Our goal is to characterize the optimal download cost $\bar{D}_L$ for fixed arbitrary $N$, $K$, and $f$. That is, to solve for
\begin{align}
     \bar{D}_L = \min \left \{ \bar{D} : (\bar{D}, L) \textrm{ is achievable} \right \}. \label{eq:DbarStar}
\end{align}
Clearly, the user can ignore its outdated message $\hat{W}_\theta$ and re-download the whole new message $W_\theta$ using standard PIR schemes \cite{SunJafar2017CapacityOfPIR}. In the next section, however, we show that we can use $\hat{W}_\theta$ to do strictly better.

%===
\section{Main Result} \label{sec:mainresult}

We present our main result in the following theorem:

\begin{theorem} \label{theorem1}
    In the private updating problem, we have 
    \begin{align} \label{eq:theorem1}
        %\bar{D}_{\bar{L}} & = \left \lceil \left \lceil \bar{L} \right \rceil \left (1 + \frac{1}{N} + \cdots + \frac{1}{N^{K-1}}  \right ) \right \rceil  = \left \lceil \frac{\left \lceil \bar{L} \right \rceil}{C} \right \rceil. 
        \left\lceil\frac{\bar{L}}{C}\right\rceil\leq\bar{D}_L\leq\left\lceil\frac{\lceil\bar{L}\rceil}{C}\right\rceil,
    \end{align}
    with $C$ and $\bar{L}$ defined in (\ref{eq:pirC}) and (\ref{eq:WbarLength}), respectively.
\end{theorem}

\begin{figure}[t]
	\centering
	\epsfig{file=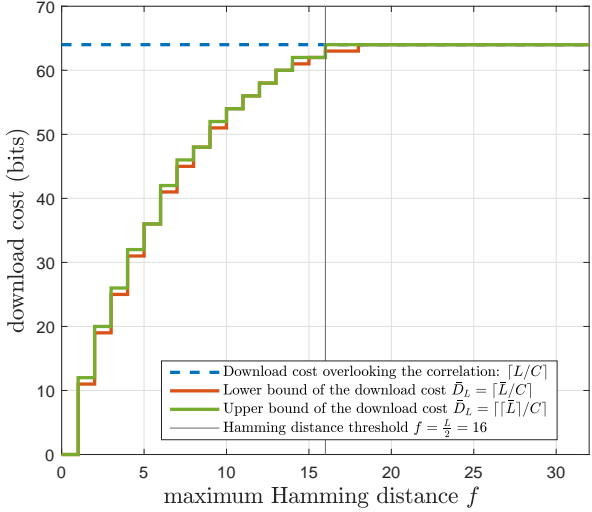,width=0.5\textwidth}
	\caption{Download cost of private updating with $L=32$ bits, $N=2$ databases, and $K=10$ messages.}
	\label{D_L_curve}
	\vspace{-.2in}
\end{figure}

Fig.~\ref{D_L_curve} shows the efficiency of our result by plotting the upper and lower bounds of the download cost for the private updating problem with $L=32$ bits, $N=2$ databases, and $K=10$ messages. 

We show the first inequality in (\ref{eq:theorem1}) by presenting a converse proof for Theorem \ref{theorem1} in Section~\ref{sec:converse}, which is based on similar arguments to those used in cache-aided PIR settings \cite{WeiBanawanUlukus2019PIRPartiallyKnownSI}. The second inequality in (\ref{eq:theorem1}) is shown by a novel achievability scheme for Theorem \ref{theorem1} in Section~\ref{sec:achieve}, which is based on distributed source coding \cite{PradhanRamchandran2003DISCUS}. We now have some remarks.

\begin{remark} \label{remark:ftheshold}
    From \eqref{eq:WbarLength} and \eqref{eq:frange}, it follows that $\left \lceil \bar{L} \right \rceil = L$ for all values of $f \geq \frac{L}{2}$; and that $\left \lceil \bar{L} \right \rceil < L$ for all values of $f < \frac{L}{2}$.\footnote{This can be readily shown using the binomial theorem. Details are omitted.} This means that there is a \emph{Hamming distance threshold} of $\frac{L}{2}$ beyond which there is no advantage to using a private updating strategy, and below which there will always be some savings in download cost (see Fig.~\ref{D_L_curve}). 
    %This may be useful if the user has some insight into the distortion of their cached message: if they suspect that at least half of the cached message symbols are incorrect, they may choose to simply re-download the entire message using a standard PIR scheme. 
    %A proof of this threshold is based on direct combinatorics, and is omitted due to space limits.
    %A proof of this threshold is provide in Section \ref{sec:appendix}.
\end{remark}

\begin{remark}
If $L$ and $f$ are such that $\bar{L}=\lceil\bar{L}\rceil$ then the two bounds in Theorem~\ref{theorem1} match. We will see that this holds if a \emph{perfect code}\footnote{Perfect codes are those that attain the Hamming bound with equality \cite{blahut2003algebraic}.} by which the queries are sent exists (cf. Section~\ref{sec:achieve}). Otherwise, if $\bar{L}<\lceil{\bar{L}}\rceil$, one can show that the two bounds are within 2 bits for $N\geq2$ databases (see \cite[Section 7.2]{SunJafar2017ArbitraryMessageLengthPIR}).
\end{remark}

% \begin{remark} \label{remark:ArbMesLenSimilar}
    % *in progress* A remark about how this result is similar to the arbitrary message length PIR paper. Also note that download cost can be improved via theorem 2 of that paper. It allows us to remove the ceiling function on $\bar{L}$ if we choose a good base to download in. Either way, the cost can be reduced by a max of 2 according to that theorem (i think). 
% \end{remark}

\section{Proof of Main Result: Converse} \label{sec:converse}

In this section, we show that $\lceil\bar{L}/C\rceil$ serves as a general lower bound for the download cost in (\ref{eq:generalDLcost}). To do so, we prove two useful lemmas, which were previously used in the cache-aided PIR setting of \cite{WeiBanawanUlukus2019PIRPartiallyKnownSI}, for the case of our private updating problem. The two lemmas are then combined to prove the general lower bound. The key difference between our lemmas and those in \cite{WeiBanawanUlukus2019PIRPartiallyKnownSI} is that rather than a set of cached messages, the user is given an outdated message $\hat{W}_\theta$, requiring careful handling of the correlation between $W_\theta$ and $\hat{W}_\theta$. Without loss of generality, we re-label the messages such that $\theta=1$.

\begin{lemma}[Interference lower bound] \label{lemma:interference}
In the private updating problem, the interference from undesired messages within the answering strings, $\bar{D} - \bar{L} $, satisfies
\begin{align}
    \bar{D} - \bar{L} \geq I\left(W_{2:K} ; Q^{[\hat{W}_1, f]}_{1:N}, A^{[\hat{W}_1, f]}_{1:N} \Big| W_1, \hat{W}_1\right). \label{eq:interference}
\end{align}
\end{lemma}

\begin{Proof}
We start with the right hand side of \eqref{eq:interference},
\begin{align}
    &I(W_{2:K} ; Q^{[\hat{W}_1, f]}_{1:N}, A^{[\hat{W}_1, f]}_{1:N} | W_1, \hat{W}_1) \nonumber \\
    & = I(W_{2:K} ; Q^{[\hat{W}_1, f]}_{1:N}, A^{[\hat{W}_1, f]}_{1:N}, W_1 | \hat{W}_1) \nonumber \\
    & \quad \quad - I(W_{2:K} ; W_1 | \hat{W}_1) \\
    & = I(W_{2:K}; Q^{[\hat{W}_1, f]}_{1:N}, A^{[\hat{W}_1, f]}_{1:N} | \hat{W}_1) \nonumber \\
    & \quad \quad + I(W_{2:K}; W_1 | Q^{[\hat{W}_1, f]}_{1:N}, A^{[\hat{W}_1, f]}_{1:N}, \hat{W}_1) \\
    & \numeq{\eqref{eq:correct}}{=} I(W_{2:K} ; Q^{[\hat{W}_1, f]}_{1:N}, A^{[\hat{W}_1, f]}_{1:N} | \hat{W}_1) \\
    & \numeq{\eqref{eq:Qindep}}{=} I(W_{2:K} ;  A^{[\hat{W}_1, f]}_{1:N} | Q^{[\hat{W}_1, f]}_{1:N}, \hat{W}_1) \\
    & = H(A^{[\hat{W}_1, f]}_{1:N} | Q^{[\hat{W}_1, f]}_{1:N}, \hat{W}_1) \nonumber \\
    & \quad \quad- H(A^{[\hat{W}_1, f]}_{1:N} | Q^{[\hat{W}_1, f]}_{1:N}, W_{2:K}, \hat{W}_1) \\
    & \numeq{\eqref{eq:correct}}{=} H(A^{[\hat{W}_1, f]}_{1:N} | Q^{[\hat{W}_1, f]}_{1:N}, \hat{W}_1) \nonumber \\
    & \quad \quad- H(A^{[\hat{W}_1, f]}_{1:N}, W_1 | Q^{[\hat{W}_1, f]}_{1:N}, W_{2:K}, \hat{W}_1) \\
    & \leq H(A^{[\hat{W}_1, f]}_{1:N} | Q^{[\hat{W}_1, f]}_{1:N}, \hat{W}_1) \nonumber \\
    & \quad \quad - H(W_1 | Q^{[\hat{W}_1, f]}_{1:N}, W_{2:K}, \hat{W}_1) \\
    & \numeq{\eqref{eq:Qindep}}{=} H(A^{[\hat{W}_1, f]}_{1:N} | Q^{[\hat{W}_1, f]}_{1:N}, \hat{W}_1) - H(W_1 | W_{2:K}, \hat{W}_1) \\
    & \numeq{\eqref{eq:generalDLcost},\eqref{eq:WbarIndep}}{\leq} \bar{D} - \bar{L}. \label{eq:InterferenceFinalStep}
\end{align}
This concludes the proof.
\end{Proof}

Note that if privacy was not a constraint, then $\bar{D} = \bar{L}$ and the interference from undesired messages would be non-existent. 
However, when the privacy constraint is present, $\bar{D} - \bar{L}$ characterizes the number of bits that will be downloaded and used as side information to preserve privacy from the databases in a given scheme. 

\begin{lemma}[Induction lemma] \label{lemma:induction}
    For all $k \in \{2, \dots, K\}$, 
    the mutual information term in Lemma~\ref{lemma:interference} can be inductively lower bounded as
    \begin{align}
        & I\left(W_{k:K} ; Q^{[\hat{W}_{k-1}, f]}_{1:N}, A^{[\hat{W}_{k-1}, f]}_{1:N} \Big| W_{1:k-1}, \hat{W}_{k-1}\right) \nonumber  \\
        & \geq \frac{1}{N} I\left(W_{k+1:K} ; Q^{[\hat{W}_{k}, f]}_{1:N}, A^{[\hat{W}_{k}, f]}_{1:N} \Big| W_{1:k}, \hat{W}_k\right) + \frac{\bar{L}}{N}. \label{eq:induction}
    \end{align}
\end{lemma}

\begin{Proof}
    We start with the left hand side of \eqref{eq:induction},
    \begin{align}
        & I(W_{k:K} ; Q^{[\hat{W}_{k-1}, f]}_{1:N}, A^{[\hat{W}_{k-1}, f]}_{1:N} | W_{1:k-1}, \hat{W}_{k-1}) \nonumber \\
        %& = \frac{N}{N} I(W_{\theta:K} ; Q^{[\hat{W}_{\theta-1}, f]}_{1:N}, A^{[\hat{W}_{\theta-1}, f]}_{1:N} | \hat{W}_{\theta-1}, W_{1:\theta-1}) \\
        & \geq \frac{1}{N} \sum^N_{n=1} I(W_{k:K} ; Q^{[\hat{W}_{k-1}, f]}_{n}, A^{[\hat{W}_{k-1}, f]}_{n} | W_{1:k-1}, \hat{W}_{k-1}) \\
        & \numeq{\eqref{eq:privacy}}{=} \frac{1}{N} \sum^N_{n=1} I(W_{k:K} ; Q^{[\hat{W}_k, f]}_{n}, A^{[\hat{W}_k, f]}_{n} | W_{1:k-1}, \hat{W}_k) \\
        & \numeq{\eqref{eq:Qindep}}{=} \frac{1}{N} \sum^N_{n=1} I(W_{k:K} ; A^{[\hat{W}_k, f]}_{n} | W_{1:k-1},\hat{W}_k, Q^{[\hat{W}_k, f]}_{n}) \\
        & \numeq{\eqref{eq:Afunc}}{=} \frac{1}{N} \sum^N_{n=1} H(A^{[\hat{W}_1, f]}_{n} |  W_{1:k-1}, \hat{W}_k, Q^{[\hat{W}_1, f]}_{n}) \\
        & \geq \frac{1}{N} \sum^N_{n=1} H(A^{[\hat{W}_1, f]}_{n} | W_{1:k-1}, \hat{W}_k, Q^{[\hat{W}_1, f]}_{1:N}, A^{[\hat{W}_1, f]}_{1:n-1}) \\
        & \numeq{\eqref{eq:Afunc}}{=} \frac{1}{N} \sum^N_{n=1} I(W_{k:K} ; A^{[\hat{W}_1, f]}_{n} | W_{1:k-1}, \hat{W}_k, Q^{[\hat{W}_1, f]}_{1:N}, A^{[\hat{W}_1, f]}_{1:n-1}) \\
        & = \frac{1}{N} I(W_{k:K} ; A^{[\hat{W}_1, f]}_{1:N} | W_{1:k-1}, \hat{W}_k, Q^{[\hat{W}_1, f]}_{1:N}) \\
        & \numeq{\eqref{eq:Qindep}}{=} \frac{1}{N} I(W_{k:K} ; Q^{[\hat{W}_1, f]}_{1:N}, A^{[\hat{W}_1, f]}_{1:N} | W_{1:k-1}, \hat{W}_k) \\
        & \numeq{\eqref{eq:correct}}{=} \frac{1}{N} I(W_{k:K} ; W_k, Q^{[\hat{W}_1, f]}_{1:N}, A^{[\hat{W}_1, f]}_{1:N} | W_{1:k-1}, \hat{W}_k) \\
        & = \frac{1}{N} I(W_{k:K} ; Q^{[\hat{W}_1, f]}_{1:N}, A^{[\hat{W}_1, f]}_{1:N} | W_{1:k}, \hat{W}_k) \nonumber \\
        & \quad \quad+ \frac{1}{N} I(W_{k:K} ; W_k | W_{1:k-1}, \hat{W}_k) \\ 
        & \numeq{\eqref{eq:WbarIndep}}{=} \frac{1}{N} I(W_{k+1:K} ; Q^{[\hat{W}_1, f]}_{1:N}, A^{[\hat{W}_1, f]}_{1:N} | W_{1:k}, \hat{W}_1) + \frac{\bar{L}}{N}. \label{eq:InductionFinalStep}
    \end{align}
    % Since the effective download cost of $\bar{L}$ is $\left \lceil \bar{L} \right \rceil$ symbols, a ceiling function is placed on the $\bar{L}$ in the second term of \eqref{eq:InductionFinalStep}. 
    This concludes the proof.
\end{Proof}

We now apply the result of Lemma~\ref{lemma:induction} recursively on that of Lemma~\ref{lemma:interference} to get the general lower bound. %This is highlighted in the next lemma, whose proof is omitted due to space limits.

\begin{lemma}
    The optimal private updating download cost satisfies the following lower bound:
    \begin{align}
        \bar{D}_L \geq \left \lceil \bar{L} \left (1 + \frac{1}{N} + \cdots + \frac{1}{N^{K-1}}  \right ) \right \rceil = \left\lceil\frac{\bar{L}}{C}\right\rceil. \label{eq:converse}
        %\bar{D}_L \geq \bar{L} \left (1 + \frac{1}{N} + \cdots + \frac{1}{N^{K-1}}  \right ) = \frac{\bar{L}}{C}. \label{eq:converse}
    \end{align}
    
\end{lemma}

\begin{Proof}
    Any private updating scheme's download cost satisfies the following series of inequalities:
    \begin{align}
        \bar{D} & \numeq{\eqref{eq:interference}}{\geq} \bar{L} + I(W_{2:K} ; Q^{[\hat{W}_1, f]}_{1:N}, A^{[\hat{W}_1, f]}_{1:N} | W_1, \hat{W}_1) \\
        & \numeq{\eqref{eq:induction}}{\geq} \bar{L} + \frac{\bar{L}}{N} + \frac{1}{N} I(W_{3:K} ; Q^{[\hat{W}_2, f]}_{1:N}, A^{[\hat{W}_2, f]}_{1:N} | W_{1:2}, \hat{W}_2) \\
        & \numeq{\eqref{eq:induction}}{\geq} \bar{L} + \frac{\bar{L}}{N} + \frac{\bar{L}}{N^2} \nonumber \\
        & \quad \quad + \frac{1}{N} I(W_{4:K} ; Q^{[\hat{W}_3, f]}_{1:N}, A^{[\hat{W}_3, f]}_{1:N} | W_{1:3}, \hat{W}_3) \\
        & \numeq{\eqref{eq:induction}}{\geq} \dots \\
        & \numeq{\eqref{eq:induction}}{\geq} \bar{L} \left (1 + \frac{1}{N} + \cdots + \frac{1}{N^{K-1}}  \right )\numeq{\eqref{eq:pirC}}{=}\frac{\bar{L}}{C}. \label{eq:lastrecursion}
    \end{align}
        Since \eqref{eq:lastrecursion} lower bounds the download cost $\bar{D}$ for {\it any} private updating scheme, it also lower bounds the download cost of the {\it optimal} private updating scheme $\bar{D}_L$. Finally, since $\bar{D}_L$ is an integer, we take the ceiling of (\ref{eq:lastrecursion}) to get (\ref{eq:converse}).
\end{Proof}
   
This shows that the first inequality of (\ref{eq:theorem1}) holds, and concludes the converse proof.

%===   
\section{Proof of the Main Result: Achievability} \label{sec:achieve}

Our achievability scheme makes use of the correlation between $W_\theta$ and $\hat{W}_\theta$ through the knowledge of their maximum Hamming distance $f$ in order to reduce the download cost. This approach is related to the problem tackled in \cite{PradhanRamchandran2003DISCUS} (without privacy constraints), in which a source is compressed given that it is correlated with some side information that is available only at the decoder. The retrieving user represents the decoder in our case, with side information $\hat{W}_\theta$. By the Slepian-Wolf coding theorem \cite{SlepianWolf1973correlatedSources}, one can noiselessly compress the source $W_\theta$ at the rate of $H(W_\theta|\hat{W}_\theta)=\bar{L}$. The {\it compressed} source is treated as a {\it new message} to be downloaded using a PIR scheme, as opposed to downloading the whole message $W_\theta$. Such scheme, however, has a message length constraint (unlike most of the PIR works in the literature). For that reason, we leverage tools from the PIR scheme with arbitrary message length in \cite{SunJafar2017ArbitraryMessageLengthPIR} to accomplish our task. The details are illustrated in the motivating examples below.

\subsection{Example: $N=2$, $K=2$, $L=3$, and $f=1$}

In this example, we have $\bar{L}=\log_2(1+3)=2$, and $C=2/3$. We show that $\bar{D}=\left\lceil\lceil\bar{L}\rceil/C\right\rceil=3$ bits is achievable. We first start by constructing a $[3,1,3]$ linear block code, which is in this case a repetition code with generator matrix $\mathsf{G}$ and parity check matrix $\mathsf{H}$ given by
\begin{align}
    \mathsf{G}=\begin{bmatrix} 1 & 1 & 1 \end{bmatrix} , \quad \mathsf{H} = \begin{bmatrix} 1 & 1 & 0 \\ 1 & 0 & 1 \end{bmatrix}.
\end{align}
Note that such code is capable of correcting at most $f=1$ error. The syndromes associated with this code are $\mathsf{s}\in\{00,01,10,11\}$. Observe that the length of $\mathsf{s}$ is exactly $\lceil\bar{L}\rceil$.

Instead of requesting $W_\theta$, the user retrieves the index of the coset in which $W_\theta$ resides in the code's standard array. That is, its corresponding syndrome
\begin{align}
    \mathsf{s}_\theta=W_\theta\mathsf{H}^T.
\end{align}
The user then compares $\hat{W}_\theta$ to all the words in that coset, and decodes $W_\theta$ as the one closest in Hamming distance. This is guaranteed to yield the unique correct message \cite{PradhanRamchandran2003DISCUS}. Therefore, the syndrome $\mathsf{s}_\theta$ efficiently represents the flipped bits' indices $\bar{W}_\theta$, and one is able to reduce the effective message length from $L=3$ to $\lceil\bar{L}\rceil=2$ by dealing with the syndrome $\mathsf{s}_\theta$ instead of $W_\theta$.

Let $W_1=[a_1,a_2,a_3]$, and $W_2=[b_1,b_2,b_3]$. The syndromes (the new messages) are given by
\begin{align}
    \mathsf{s}_1&=W_1\mathsf{H}^T=\begin{bmatrix}a_1+a_2 & a_1+a_3\end{bmatrix}\triangleq\begin{bmatrix}\bar{a}_1 & \bar{a}_2\end{bmatrix}, \\
    \mathsf{s}_2&=W_2\mathsf{H}^T=\begin{bmatrix}b_1+b_2 & b_1+b_3\end{bmatrix}\triangleq\begin{bmatrix}\bar{b}_1 & \bar{b}_2\end{bmatrix}.
\end{align}
Assume $\theta=1$. Since $\lceil\bar{L}\rceil=N^{K-1}$, we can apply a {\it non-symmetric} PIR scheme as follows to decode $\mathsf{s}_1$ \cite{SunJafar2017ArbitraryMessageLengthPIR}:

%\begin{center}
%\begin{tabular}{ |c|c| } 
% \hline
% Database 1 & Database 2 \\ 
% \hline
% $\bar{a}_1$, $\bar{b}_1$ & $\bar{a}_2 + \bar{b}_1$ \\ 
% \hline
%\end{tabular}
%\end{center}

\begin{table}[h!]
\centering
\begin{tabular}{|M{2.5cm}|M{2.5cm}|}
\hline
Database $1$ & Database $2$ \\ \hline
\vspace{.025in} $\bar{a}_1$, $\bar{b}_1$ & \vspace{.025in} $\bar{a}_2 + \bar{b}_1$ \\ \hline
\end{tabular}
\end{table}

\noindent This has a download cost of $\bar{D}=3$ bits, which is optimal in this case since it meets the converse bound.

The repetition code used in this example is a {\it perfect code.} While this makes $\bar{L}$ an integer, and meets the converse bound, perfect codes are scarce. In the next example, we show how the proposed scheme performs with non-perfect codes.

\subsection{Example: N=2, K=2, L=5, and f=1}

In this example, we have $\bar{L}=\log_2(1+5)=2.58$, and $C=2/3$. We show that $\bar{D}=\left\lceil\lceil\bar{L}\rceil/C\right\rceil=5$ bits is achievable. As in the previous example, we start by constructing a $[5,2,3]$ linear block code. Differently though, this is not a repetition code, and is characterized by
\begin{align}
    \mathsf{G}=\begin{bmatrix} 1 & 0 & 1 & 1 & 1 \\ 0 & 1 & 1 & 1 & 0 \end{bmatrix} , \quad \mathsf{H} = \begin{bmatrix} 1 & 1 & 1 & 0 & 0 \\ 1 & 1 & 0 & 1 & 0 \\ 0 & 1 & 0 & 0 & 1 \end{bmatrix}.
\end{align}
The syndromes $\mathsf{s}$ have length $\lceil\bar{L}\rceil$. Specifically,
\begin{align}
    \mathsf{s}_1&=W_1\mathsf{H}^T=\begin{bmatrix}a_1+a_2+a_3 & a_1+a_2+a_4 & a_2+a_5\end{bmatrix} \nonumber \\
    &\triangleq\begin{bmatrix}\bar{a}_1 & \bar{a}_2 & \bar{a}_3\end{bmatrix}, \\
    \mathsf{s}_2&=W_2\mathsf{H}^T=\begin{bmatrix}b_1+b_2+b_3 & b_1+b_2+b_4 & b_2+b_5\end{bmatrix} \nonumber \\ 
    &\triangleq\begin{bmatrix}\bar{b}_1 & \bar{b}_2 & \bar{b}_3\end{bmatrix}.
\end{align}

Since $\lceil\bar{L}\rceil=N^{K-1}+1$, we follow the methodology in \cite{SunJafar2017ArbitraryMessageLengthPIR}; we privately download $N^{K-1}=2$ bits ($\bar{a}_1$ and $\bar{a}_2$) using the non-symmetric PIR scheme in the previous example, and then privately download the remaining $1$ bit ($\bar{a}_3$) using the scheme in \cite{ShahRashmiRamchandran2014OneExtraBitPIR}. The technique in \cite{ShahRashmiRamchandran2014OneExtraBitPIR} in this case is such that the user requests random linear combinations of $[\bar{a}_3~\bar{b}_3]$ from database $1$ using a random binary vector ${\bm h}$, and the same from database $2$ yet with ${\bm h}^\prime={\bm h}+{\bm e}_\theta$, where ${\bm e}_i$ is the $i$th standard basis vector. The full PIR scheme is as follows:

\begin{table}[h!]
\centering
\begin{tabular}{|M{2.5cm}|M{2.5cm}|}
\hline
Database $1$ & Database $2$ \\ \hline
\vspace{.025in} $\bar{a}_1$, $\bar{b}_1$ & \vspace{.025in} $\bar{a}_2 + \bar{b}_1$ \\ \hline
\vspace{.025in} $h_1\bar{a}_3+h_2\bar{b}_3$ & \vspace{.025in} $(h_1+1)\bar{a}_3+h_2\bar{b}_3$ \\ \hline
\end{tabular}
\end{table}

\noindent This has a download cost of $\bar{D}=5$ bits, which is 1 bit away from the converse bound since the code used is non-perfect.

\subsection{The General Scheme}

For general $N$, $K$, $L$, and $f$, we construct an $[L,L-\lceil\bar{L}\rceil,2f+1]$ linear block code. From the Gilbert-Varshamov bound \cite{blahut2003algebraic}, we know that such a code exists if
\begin{align}
    2^{\lceil \bar{L} \rceil} \leq \sum_{j=0}^{2f} {L \choose j}. \label{eq:GVbound}
\end{align}
In addition, such a code must satisfy the Hamming bound \cite{blahut2003algebraic}:
\begin{align}
    \sum_{j=0}^{f} {L \choose j} \leq 2^{\lceil \bar{L} \rceil}. \label{eq:HammingBound}
\end{align}
By the definition of $\bar{L}$ in \eqref{eq:WbarLength}, both (\ref{eq:GVbound}) and (\ref{eq:HammingBound}) are satisfied, and so the code exists and is able to correct $f$ bit flips.

Next, we map each message to its corresponding syndrome of the constructed code, which is of length $L-(L-\lceil\bar{L}\rceil)=\lceil\bar{L}\rceil$. The user then retrieves the syndrome $\mathsf{s}_\theta$ according to a PIR scheme with $N$ databases, $K$ messages, and $\lceil\bar{L}\rceil$ message length. By \cite[Theorem 1]{SunJafar2017ArbitraryMessageLengthPIR}, a download cost of $\left\lceil\lceil\bar{L}\rceil/C\right\rceil$ is achievable in this case. Finally, correctness is guaranteed since querying for the syndrome $\mathsf{s}_\theta$ allows the user to decode $W_\theta$ as the unique word in the syndrome's coset with the least Hamming distance from $\hat{W}_\theta$ \cite{PradhanRamchandran2003DISCUS}. 

This shows that the second inequality of (\ref{eq:theorem1}) holds, and concludes the achievability proof.

\section{Conclusions and Discussions}

In this work, a novel private updating problem has been introduced, in which a user's outdated message is to be privately updated by querying a set of replicated and non-colluding databases that have the up-to-date version. Under a Hamming distortion measure between the outdated and the up-to-date messages, a syndrome decoding technique is leveraged to compress the number of bits that needs to be downloaded in order to correctly update the message. This has been combined with PIR schemes with message length constraints to guarantee privacy. The proposed private updating scheme has been shown to be optimal when the system parameters enable the construction of a perfect code according to which the syndrome decoding technique is worked out. In other cases, the achievable download cost has been shown to be within at most $2$ bits from a derived converse bound.

% \subsection{The Case with Known Distortion}

The model of this paper assumes that the Hamming distortion between $W_\theta$ and $\hat{W}_\theta$ is upper bounded by $f$. If, instead, the Hamming distortion is {\it known  to be exactly} $f^\prime$, then the download cost can be reduced, and using codes to map the messages into syndromes may be unnecessary. To see this, consider an example with $N=2$, $K=2$, $L=8$, and $f=1$. In this case, $\lceil\bar{L}\rceil=4$, and by Theorem~\ref{theorem1}, a download cost of $6$ bits is achievable.% (which is $1$ bit from the converse bound). 

Let us now set $f^\prime=1$, i.e., the user knows that $W_\theta$ and $\hat{W}_\theta$ differ in {\it exactly} $1$ bit. Assuming $\theta=1$, define $\bar{a}_1\triangleq a_1+a_2+a_3+a_4$, $\bar{a}_2\triangleq a_1+a_2+a_5+a_6$, and $\bar{a}_3\triangleq a_1+a_3+a_5+a_7$, where $a_i$'s represent the bits of the desired message $W_1$. The user then constitutes similar combinations using the bits of the outdated message $\hat{W}_1$ to get $\hat{a}_1$, $\hat{a}_2$, and $\hat{a}_3$. Now observe that possessing the {\it new message} $[\bar{a}_1,\bar{a}_2,\bar{a}_3]$ is sufficient to determine the position of the flipped bit by comparing it to $[\hat{a}_1,\hat{a}_2,\hat{a}_3]$. For instance, if $\bar{a}_i=\hat{a}_i,~\forall i$, then $\hat{W}_1(8)$ needs to be flipped. If on the other hand $\bar{a}_i\neq\hat{a}_i,~\forall i$, then $\hat{W}_1(1)$ needs to be flipped. While if $\bar{a}_1\neq\hat{a}_1$, and $\bar{a}_i=\hat{a}_i,~i=2,3$, then $\hat{W}_1(4)$ needs to be flipped, and so on. Therefore, the effective message length is reduced to $3$ (as opposed to $\lceil\bar{L}\rceil=4$), and a download cost of $5$ bits is achievable.

The above procedure can be done using a {\it bisection search} approach. The user can first retrieve $a_1+a_2+a_3+a_4$, and compares it to the sum of the outdated message's first $4$ bits. If they are equal, then the error must lie in the last $4$ bits of $\hat{W}_1$. Assuming this is the case, the user downloads $a_5+a_6$, and compares it to $\hat{W}_1(5)+\hat{W}_1(6)$. If they too are equal, then the error must lie in the last $2$ bits of $\hat{W}_1$. Assuming this is the case as well, the user finally downloads $a_7$ and compares it to $\hat{W}_1(7)$. If they are equal, then $\hat{W}_1(8)$ needs to be flipped. We see that this bisection approach has an effective message length of $\log_2(L)=3$ bits. However, since the next query structure depends on the answers of the previous queries, a {\it multiround} PIR scheme needs to devised in this case \cite{sun2018multiround}.

It would be interesting to extend the results of this paper to work for the case of known distortion $f^\prime$ (and generally for other notions of correlation measures between $W_\theta$ and $\hat{W}_\theta$), which may be relevant in certain applications.

\end{document}